\documentclass[twocolumn,aps,prb,amssymb,amsmath,floatfix,showpacs]{revtex4}
\usepackage{graphicx,subfigure}
\usepackage{bm}
\usepackage{verbatim}
\usepackage{amsmath}
\usepackage{amssymb}
\usepackage[T1]{fontenc}
\usepackage{ae,aecompl}

\newcommand{\re}{{(r)}}

\newcommand{\rv}{{\bf r}}

\newcommand{\xv}{{\bf x}}

\newcommand{\qv}{{\bf q}}

\newcommand{\grad}{{\bm{\nabla}}}

\newcommand{\uv}{{\bf u}}

\newcommand{\hR}{{\hat R}}
\newcommand{\xh}{{\hat{\bf x}}}
\newcommand{\yh}{{\hat{\bf y}}}
\newcommand{\zh}{{\hat{\bf z}}}

\newcommand{\oh}{{\frac{1}{2}}}

\def\rf#1{(\ref{#1})}
\def\rfs#1{Eq.~\rf{#1}}

\begin{document}
\title{Smectic vortex glass}
\author{Leo Radzihovsky}
\affiliation{Department of Physics and Center for Theory of Quantum Matter\\
University of Colorado, Boulder, CO 80309}
\date{\today}
\email{radzihov@colorado.edu}
\date{\today}

\begin{abstract}
  We show that in type-II superconductors a magnetic field applied
  transversely to correlated columnar disorder, drives a phase
  transition to a distinct ``smectic'' vortex glass (SmVG) state. SmVG
  is characterized by an infinitely anisotropic electrical transport,
  resistive (dissipationless) for current perpendicular to (along)
  columnar defects. Its positional order is also quite unusual,
  long-ranged with true Bragg peaks along columnar defects and
  logarithmically rough vortex lattice distortions with quasi-Bragg
  peaks transverse to columnar defects. For low temperatures and
  sufficiently weak columnar-only disorder, SmVG is a true
  topologically-ordered ``Bragg glass'', characterized by a vanishing
  dislocation density. At sufficiently long scales the residual
  ever-present point disorder converts this state to a more standard,
  but highly anisotropic vortex glass.
\end{abstract}


\maketitle
\section{Introduction}
\label{introduction}

\subsection{Background and motivation}
\label{background}

The discovery of high-temperature superconductors, now more than 30
years ago, in parallel with a search for the microscopic
``mechanism'', (that continues todate) generated vigorous studies of
vortex states of matter in the presence of thermal fluctuations,
pinning disorder and electrical (``super''-) current in and out of
equilibrium\cite{BlatterRMP,NattermannScheidl,FFH,HR,LeDoussalBCSchapter,
drivenRefs}, predicting and finding a rich magnetic field ($H$) -
temperature ($T$) phase diagram of these type-II
superconductors\cite{Tinkham}.

In contrast to a mean-field picture, thermal fluctuations drive a
first-order melting of a vortex lattice over a large portion of the
phase diagram into a resistive (though highly diamagnetic) vortex
liquid\cite{Eilenberger,DSFisher,NelsonSeung,meltingExp}.

In the low-temperature vortex solid state, arbitrarily weak point
pinning disorder, on sufficiently long scale\cite{Larkin,ImryMa}
always disrupts translational order of the vortex lattice. Supported
by experimental observations\cite{Koch}, it was argued
\cite{MatthewFisherVG,CO, TonerDiVincenzo,FFH}, that the resulting
vortex glass state is characterized by an Edwards-Anderson\cite{EA}
order parameter, with vortices collectively pinned, and thereby
exhibiting a vanishing linear mobility, implying a zero linear
resistivity of the vortex glass state.\cite{FFH} For weak disorder, a
distinct topologically-ordered vortex Bragg glass, characterized by a
vanishing density of unpaired dislocations and concomitant power-law
decay of crystalline order, was also proposed\cite{GLbragg}, supported by further
analytical\cite{stabilityCarpentierBraggGlass1996,stabilityKierfeldBraggGlass1997,
  DSFisherBG}, numerical\cite{GingrasHuse1996,vanOtterlo1998} analyses
and by neutron scattering experiments\cite{KleinNatureBragg2001}.

Introduction (via heavy ion irradiation) of columnar pinning defects
significantly enhances pinning\cite{columnarExp}, and for a magnetic
field along columnar defects, was predicted to lead to an {\em
  anisotropic} vortex glass dubbed ``Bose
glass''\cite{FisherLee,NelsonVinokur} because of its mathematical
connection to interacting two-dimensional (2D) quantum bosons pinned
by a quenched (time-independent) random 2D potential\cite{BoseGlass}.

\begin{figure}[tbp]
\includegraphics[width=2.2in]{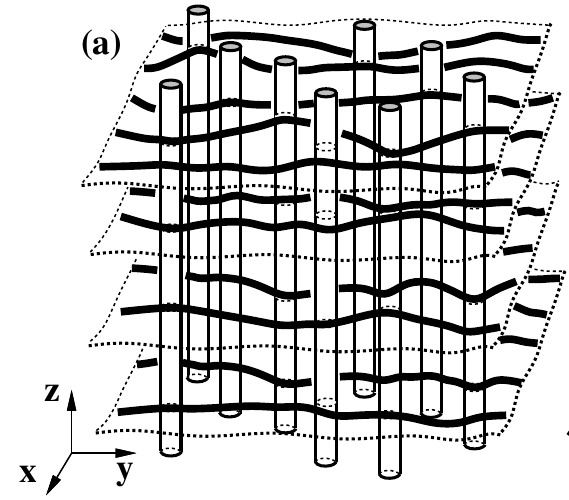}
\includegraphics[width=1in,trim=0 0 0 -1cm]{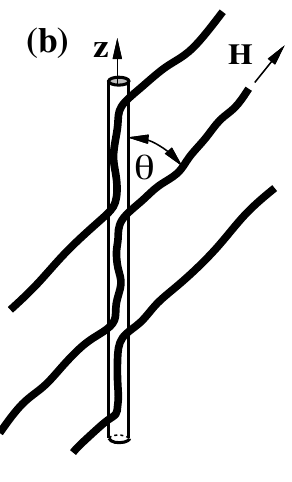}
\includegraphics[width=2.5in]{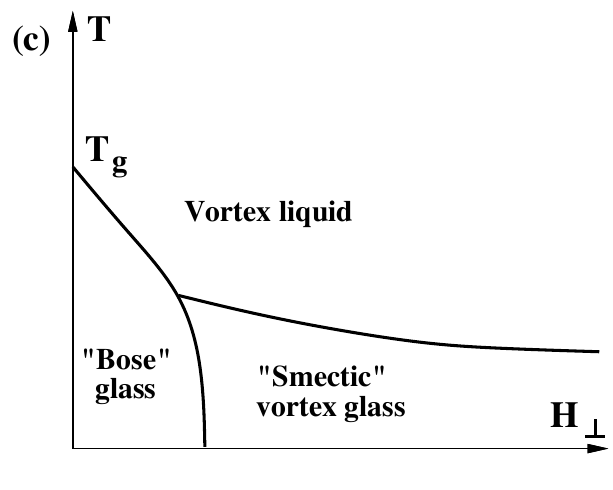}
\put (-128,0) {$H_\perp^{c_1}$}
\caption{(a) Illustration of the {\em transverse} magnetic field
  geometry in a positionally random array of parallel columnar
  defects, leading to the ``smectic vortex glass'' (SmVG) phase in
  (c). (b) Vortex lines' tilt response to an applied field, applied
  transversely to a columnar defect. (c) A schematic phase diagram,
  illustrating a transverse Meissner effect up to field
  $H_\perp^{c_1}(T)$, at which at low $T$ a fully superconducting
  ``Bose glass'' undergoes a transition to a SmVG.}
\label{fig:SmVGfig}
\end{figure}

One key feature of the vortex Bose glass, that qualitatively
distinguishes it from the corresponding isotropic vortex glass is the
existence of the ``transverse'' Meissner effect,\cite{NelsonVinokur},
namely a vanishing response to a field $H_\perp < H_\perp^{c_1}$,
applied transversely to columnar defects.  This expulsion of the
transverse flux density, $B_\perp$, that has received considerable
experimental\cite{trMeissnerExp} and simulations\cite{Wallin} support,
corresponds to an effectively divergent tilt
modulus\cite{NelsonVinokur,BalentsEPL} inside this anisotropic vortex
glass, that in the quantum correspondence maps onto a vanishing
superfluid density in the Bose glass phase. A detailed theoretical
description of the transverse Meissner effect (as well as other
properties of the phase) has been predominantly limited to
noninteracting vortex lines\cite{NelsonVinokur,Hwa}, supported by
variable-range hopping\cite{EfrosShklovskii} scaling theories
\cite{NelsonVinokur,Hwa,NelsonRadzihovsky,Wallin}, analysis in reduced
planar geometry\cite{Affleck,LRtiltPRB} and
simulations\cite{Wallin,TauberNelson}.

As illustrated in Fig.\ref{fig:SmVGfig}(c), vortex Bose glass is thus
confined to the low-temperature and low-transverse magnetic field part
of the phase diagram\cite{NelsonVinokur}. To date, it has been tacitly
assumed that beyond the critical value $H_\perp^{c_1}$ of the
transverse field, the tilted state is a vortex liquid or crystal (a
conventional vortex glass in the presence of risidual point
disorder\cite{Hwa}) with a finite resistivity and a finite tilt
modulus, qualitatively the same vortex phase appearing above melting
transition at $T_g$. Our goal here is to explore and characterize a
highly tilted vortex geometry illustrated in Fig.\ref{fig:SmVGfig}(a),
which, in the absence of point disorder at large tilting angle
($\sim\pi/2$), $H_\perp \gg H_\perp^{c_1}$ we show is a qualitatively
distinct vortex glass phase, that we dub as ``smectic'' vortex glass
(SmVG).  In fact, dating almost 20 years back, this regime has been
explored experimentally\cite{JaegerPRL2000PRB2001}, finding
interesting anomalous behavior in the magnetic ac susceptometry and
electrical transport, and motivating our theoretical study, that has
taken this long to formalize.

The rest of the paper is organized as follows. We conclude the
Introduction with a summary of our main results. In Sec. II we
introduce an elastic model to describe a vortex array, tilted at a
large angle relative to the random forest of parallel columnar
defects.  Because pinning disorder does not couple to displacements
along columnar defects, in Sec. III we reduce the system to an
effective scalar model for a transverse-only distortions characterized
by long-range elasticity, resembling an effective planar vortex glass
model\cite{MatthewFisherVG}.  We analyze this model within and beyond
the Larkin length scale\cite{Larkin,ImryMa} and thereby predict the
existence of a new glassy state of anisotropic vortex matter, that
because of its periodic positional order along columnar defects we
refer to as ``smectic'' vortex glass. In Sec. IV we discuss the
stability of the SmVG to dislocations and conclude in Sec. V with the
summary, discussion and open questions.

\subsection{Summary of the results}

As illustrated in Fig.\ref{fig:SmVGfig}a, we focus on the large tilt
angle, around a configuration of vortex lines transverse to columnar
defects, staying away from the nontrivial vicinity of the critical
field $H_\perp^{c_1}$.\cite{Hwa,LRtiltPRB}

\begin{figure}[tbp]
\includegraphics[width=3in]{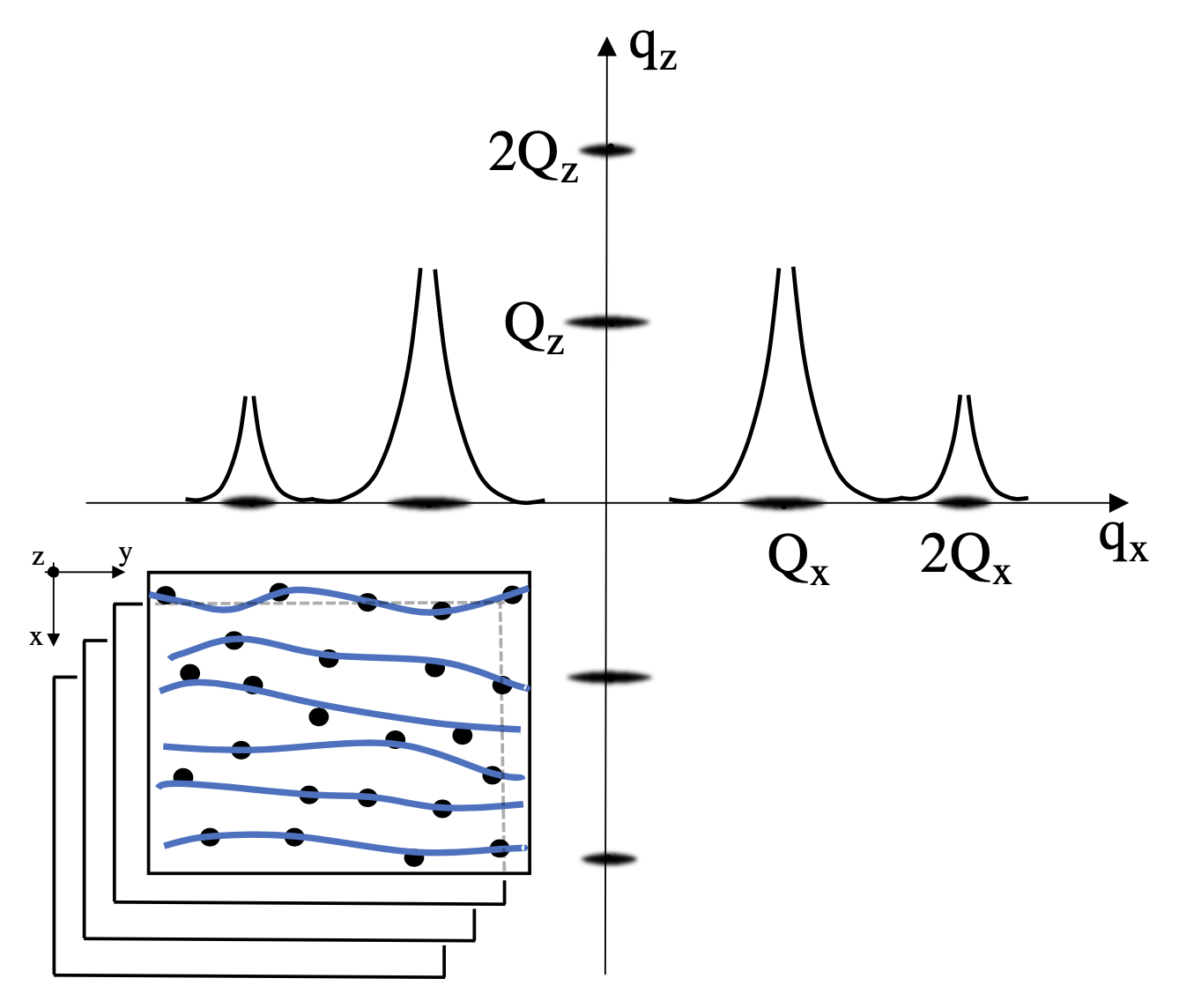}
\caption{A schematic form of the structure function $S(q_x,q_z)$
  characterizing smectic vortex glass order. In the absence of point
  disorder it exhibits resolution-limited $\delta$-function Bragg
  peaks along columnar defects ($q_z$) and universal quasi-Bragg peaks
  along the transverse axis ($q_x$).}
\label{StructureFncSVGfig}
\end{figure}

In this geometry, in the absence of point disorder, we predict that a
qualitatively distinct, highly anisotropic ``smectic'' vortex
glass\cite{commentSm} emerges. Structurally, it is characterized by a
periodic vortex order {\em along} the translationally-invariant
columnar defects axis with smectic-like (but here $\delta$-function
``true'') Bragg peaks
\begin{equation}
  S(0,q_z)\sim \sum_{n}I_{n Q_z}\delta(q_z - n Q_z)
  \label{SqzBragg}
\end{equation}
illustrated in Fig.\ref{StructureFncSVGfig}, with $I_Q(T)$ a standard
phonon Debye-Waller factor.

In contrast, vortices are randomly pinned transversely by
cross-sections of columnar defects (with each plane akin to a
Cardy-Ostlund planar vortex glass\cite{CO,MatthewFisherVG,
  TonerDiVincenzo}), with ``rough'' transverse distortions, that for
weak disorder are logarithmically (rather than $\ln^2L)$ rough,
\begin{equation}
\oh\overline{\langle u(x)-u(0)\rangle^2}\approx Q_x^{-2} \eta\ln(x/a), 
\end{equation}
with a universal amplitude $\eta\approx 2\pi^2/9$ in 3D, (identical to
that found in Ref.\onlinecite{GLbragg}) for lengths exceeding the
Larkin scale,
\begin{equation}
 \xi_{\text{L}} = a\sqrt{\tilde K^2/\Delta},
 \label{xiLarkinIntro}
\end{equation}
where $\tilde K$ is the effective elastic constant and $\Delta$ the variance
of the random force pinning disorder.

We thus predict that SmVG displays quasi-Bragg peaks\cite{GLbragg},
transverse to columnar defects,
\begin{equation}
  S(q_x,0) \sim \sum_n\frac{1}{|q_x - n Q_x|^{1-n^2\eta}},
\end{equation}
illustrated in Fig.\ref{StructureFncSVGfig}, with a {\em universal}
power-law exponent $\eta\approx 2\pi^2/9$.  We note that if this
approximate value of $\eta$ is indeed greater than $1$, we then
predict only cusp singularities, rather than divergent quasi-Bragg
peaks, to appear at the reciprocal lattice vectors, $n Q_x$,
$n\in \mathbb{Z}$.

To the extent that the disorder imposed by columnar defects is
perfectly correlated along their axis (taken as $z$), we predict that
vortex planes of the SmVG, transverse to columnar defects exhibit true
2D topological ``Bragg'' glass
order\cite{MatthewFisherVG,DSFisherFRG,GLbragg,
  KleinNatureBragg2001,DSFisherBG,RTaerogel}, characterized by a
vanishing dislocation density, with only elastic single-valued
distortions as illustrated in Fig.\ref{fig:SmVGfig}a and inset of
Fig. \ref{StructureFncSVGfig}.

For strong disorder and/or high temperature, dislocations may
spontaneously proliferate, driving a topological transition to a fully
disordered vortex glass-like state\cite{FFH}, with transverse
distortions characterized by a Lorentzian structure function along
$q_x$.

\begin{figure}[tbp]
\includegraphics[width=3in]{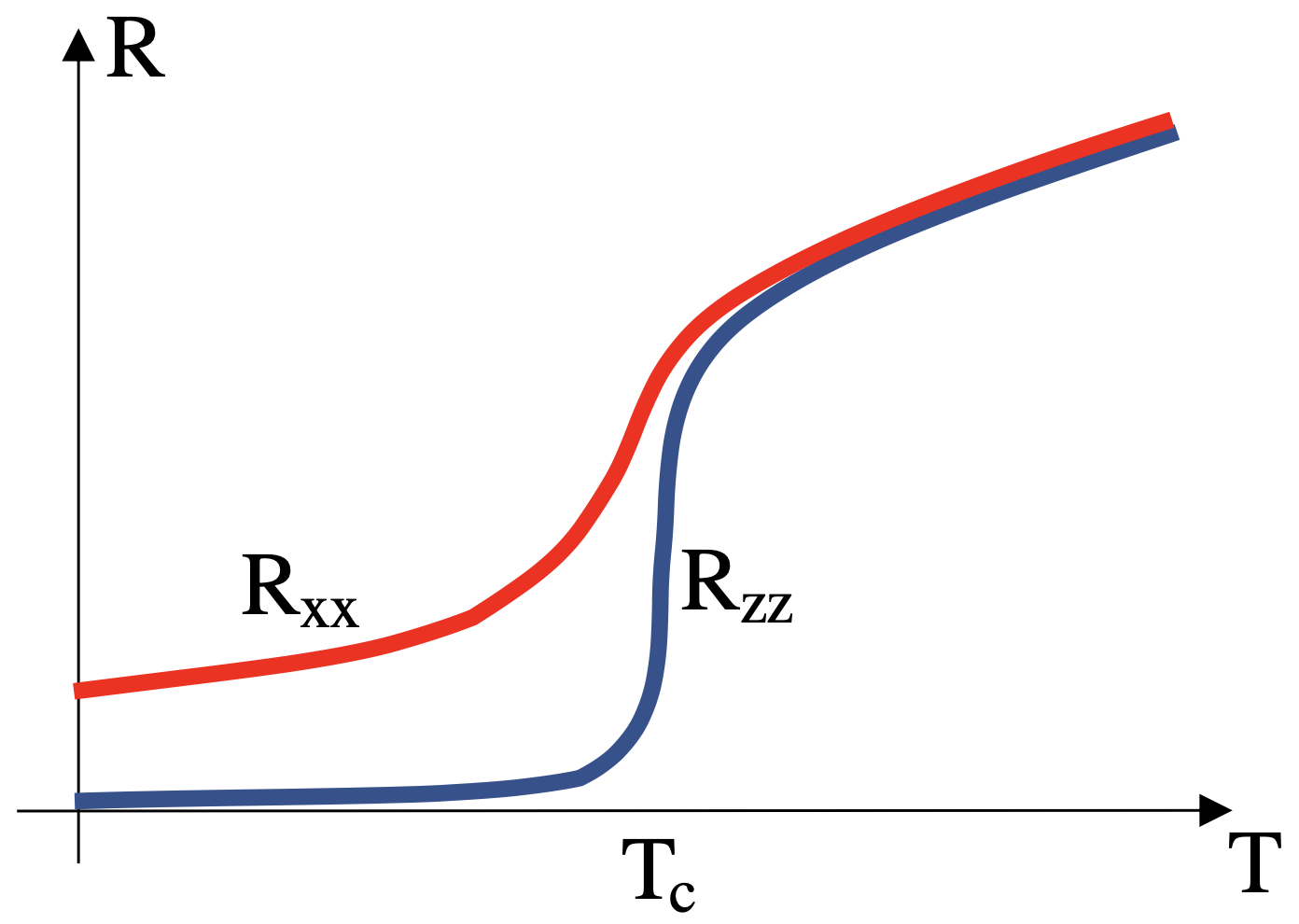}
\caption{A schematic illustration of infinitely anisotropic resistance
  (in the absence of additional point disorder) along $R_{zz}$ and
  perpendicular $R_{xx}$ to columnar defects characterizing the
  smectic vortex glass state. In a more realistic case of residual
  point pinning, at sufficiently low temperatures and long scales the
  state will freeze into a fully dissipationless superconducting
  highly anisotropic vortex glass.}
\label{ResistanceAnisotropyFig}
\end{figure}

Concomitantly, as illustrated in Fig.\ref{ResistanceAnisotropyFig}
we predict such smectic vortex glass state (in the idealized case of
no additional pinning disorder) to exhibit a striking infinite
resistive anisotropy. For current transverse to columnar defects and
to the magnetic field, a free motion of vortices along homogenous
columnar defects leads to a nonzero flux-flow resistivity
$\rho_{\perp}\sim (B/H_{c2})\rho_n$, with $\rho_n$ resistivity arising
from normal carriers residing in the vortex core, superfluid phase
slip and time-dependent flux associated with moving vortices.  In contrast, for
current along columnar defects vortex motion is impeded by columnar
defects, exhibiting a vanishing linear resistivity of the vortex glass
state\cite{MatthewFisherVG,FFH}. Thus, we predict a vanishing of the
resistivity ratio
\begin{equation}
\rho_{zz}/\rho_{xx}\rightarrow 0,
\end{equation}
inside the smectic vortex glass.

We also predict that the smectic vortex glass, viewed as an effective
elastic medium exhibits a vanishing response $\partial_z u_x$ to shear
stress $\sigma_{zx}$ in the $zx$ plane transverse to applied magnetic
field (i.e., transverse to vortex lines), defined by the columnar
defects ($z$) axis. SmVG is thus characterized by a divergent shear
modulus
\begin{equation}
  \mu_{zx}\rightarrow\infty.
  \label{muzx}
\end{equation}

However, unlike Bose glass vortex state, where such divergent tilt
modulus can be readily probed via a vanishing response to a magnetic
field transverse to columnar defects - ``transverse Meissner effect''
\cite{NelsonVinokur,Hwa,BalentsEPL}, here there does not appear to be
a simple way to probe the vanishing SmVG response corresponding to
\rf{muzx}.


We now turn to the vortex model in this transverse geometry and to the
derivation of these results.

\section{Model of the smectic vortex glass}
\label{model}
In a type-II superconductor, for fields above a lower-critical field,
$H_{c1}$ the magnetic flux penetrates in a form of interacting vortex
flux lines carrying a unit of a fundamental flux quantum
$\phi_0=hc/2e$, with average density determined by the applied
magnetic field.\cite{Tinkham} At low temperature and in the absence of
disorder, repulsive directed vortex lines crystalize into a periodic
triangular array - Abrikosov vortex lattice, whose elastic
description\cite{elasticity,GLelastic,BlatterRMP} can be derived from
the Ginzburg-Landau theory for the superconducting order parameter,
that itself, under certain conditions, is derivable from the
microscopic theory.

\subsection{Elasticity and columnar pinning for a transverse magnetic field}

Transcending such a detailed derivation, on sufficiently long length
scale the elastic vortex lattice energy functional can be deduced
purely on symmetry grounds. In three-dimensions, it is formulated in
terms of a two-dimensional Eulerian phonon vector field (Goldstone
modes of the spontaneously broken translational symmetry)
${\bf u}({\bf r}_\perp,y)=\left(u_x(x,y,z), u_z(x,y,z)\right)$
describing vortex lattice distortion in the $x-z$ plane (spanned by
${\bf r}_\perp=(x,0,z)$), transverse to the vortex lines, that we take
to run along the $y$-axis, as illustrated in Fig.\ref{fig:SmVGfig}(a).

Such a triangular vortex crystal state is then characterized by an
elastic energy, that, in the absence of other ingredients, is an
isotropic functional of $\bf u$,
\begin{eqnarray}
H_{el} = \int_\rv \left[\frac{K}{2}(\partial_y{\bf u})^2 + \mu
  u_{ij}^2 + \frac{\lambda}{2} u_{ii}^2\right],
\label{Hel}
\end{eqnarray}
with $\int_\rv\equiv\int d^2r_\perp dy\equiv\int d\xv dz$,
$\xv = (x,y,0)$, $K$ a tilt modulus and $\mu,\lambda$ the elastic
Lam\'e coefficients.\cite{commentLatticeAnisotropy}

A random array of parallel columnar defects (see
Fig.\ref{fig:SmVGfig}) oriented along $\zh$, transversely to vortex
lines along $\yh$ introduces a random highly anisotropic pinning
potential $V(x,y)$, that couples to the two-dimensional (in a plane
spanned by ${\rv_\perp} = (x,0,z)$) vortex density
\begin{equation}
n({\rv})\approx n_{0}-n_0{\grad}_\perp\cdot{\bf u} + \sum_{\bf Q} n_{\bf
  Q}\;e^{i {\bf Q}\cdot({\rv}_\perp+{\bf u}(\rv))},
\label{nv}
\end{equation}
and a random tilting potentials $\delta K_i(x,y)$, that for the
$\pi/2$ transverse-field geometry of Fig.\ref{fig:SmVGfig}(a) couples
to the {\em even} power of the vortex lattice tilt $\partial_y{\bf
  u}$, thereby preserving the $\pm \partial_y{\bf u}$ symmetry. The
resulting pinning energy functional is given
by\cite{commentRandomStress}
\begin{eqnarray}
H_{pin} &=& \int_\rv\left[
\frac{\delta K_i(x,y)}{2}(\partial_y u_i)^2+
n_0 V(x,y){\grad}_\perp\cdot{\bf u}\right.\nonumber\\
&&\left.-V(x,y)\sum_{\bf Q} n_{\bf
  Q}\;e^{i {\bf Q}\cdot({\rv}_\perp+{\bf u}(\rv))}\right],
\label{Hpin}
\end{eqnarray}
where $n_0$ is the average density and $n_{\bf Q}$ are the Fourier
components of the vortex density at the discrete set of reciprocal
lattice vectors, ${\bf Q}$.
By symmetry, the average columnar defects density results in a biaxial
lattice anisotropy\cite{commentLatticeAnisotropy} and the tilt
modulus, with correction $\delta K_i = \overline{\delta K_i(x,y)}$,
expected to be $\delta K_z < 0$ and $\delta K_x\approx 0$. This can be
accounted for by $K\rightarrow K_i$. Because $\delta K_i(x,y)$ is
coupled to the square of the vortex tilt, for weak heterogeneity
fluctuations around its average are subdominant at long scales.

Furthermore, because columnar defects are translationally invariant,
the corresponding pinning potential $V(x,y)$ is $z$-independent, and
thus long-wavelength pinning
$\int_{\rv}n_0 V(x,y){\grad}_\perp\cdot{\bf u}$ reduces to
$\int_{\rv} U_0(x,y)\partial_x u_x$, only coupling to the $u_x(x,y)$
displacement transverse to the columnar defect. As in other random
pinning problems\cite{DSFisherFRG,GLbragg,RTaerogel} standard analysis
shows that the long-wavelength part of the pinning is perturbatively
subdominant for weak disorder in 3D (in contrast to the 2D
Fisher-Cardy-Ostlund pinning, where it leads to a super-rough
glass\cite{MatthewFisherVG,CO, TonerDiVincenzo} with $\ln^2r$
correlations). However, as we will see it does play a role at
asymptotic scales in the physical 3D case.

The $z$ independence of $V(x,y)$ also reflects itself in the averaging
out of the short-scale density components $n_{\bf Q}$ with a nonzero
${Q_z\zh}$ reciprocal lattice vector.  Thus, $u_z(\rv)$ only appears
harmonically, and the nonlinear short-scale pinning only acts on
$u_x(\rv)$. The full effective Hamiltonian then reduces
to\cite{commentRandomStress}
\begin{eqnarray}
H &=& \int_\rv\left[\oh u_i\hat{\Gamma}_{ij} u_j 
+ U_0(x,y)\partial_x u_x + U(x,y, u_x(\rv))\right],\;\;\;\;\;\; 
\label{Hfinal}
\end{eqnarray}
where we take $U_0(x,y)$ to be characterized by a zero-mean Gaussian
distribution with variance $R_0$
\begin{equation}
\overline{U_0(x,y)U_0(x',y')}=R_0\delta(x - x')\delta(y - y').
\end{equation}
The short-scale pinning potential 
\begin{equation}
U(x,y, u_x(\rv))=\sum_{Q} U_Q(x,y) e^{i Q u_x(\rv)},
\end{equation}
has Fourier component at $Q\equiv Q_x$ given by
\begin{equation}
  U_Q(x,y)=-\int_{\in
    \text{unit-cell at}\; x} d\delta x V(x+\delta x,y) n_Q e^{i Q\delta
    x},
\end{equation}
and we take it to be zero-mean, Gaussian correlated and characterized
by a variance
\begin{equation}
\overline{U(x,y, u_x)U(x',y', u_x')}=R(u_x - u_x')\delta(x -
x')\delta(y - y').
\label{Rdefine}
\end{equation}
The periodicity of $U(x,y, u_x+2\pi/Q) = U(x, y, u_x)$ (and therefore
of $R(u_x)$) in $u_x$ reflects the identity symmetry of vortex
lines. The Fourier transform of the inverse propagator
$\hat{\Gamma}_{ij}$ [for now ignoring (here) unimportant
anisotropies\cite{commentLatticeAnisotropy}] is given by
\begin{equation}
\Gamma_{ij}(\qv) = \left(K q_y^2 + \mu q_\perp^2\right)P^T_{ij} +
\left(K q_y^2 + (2\mu+\lambda) q_\perp^2\right)P^L_{ij},
\end{equation}
with $P^{T,L}_{ij}(\qv_\perp)$ the transverse and longitudinal
projection operators with respect to $\qv_\perp$.

\subsection{Reduction to a correlated random-field xy model}

It is now quite clear how to proceed. Because $u_z$ only enters the
Hamiltonian harmonically, we can integrate it out exactly, obtaining
an effective xy-model for a single phonon $u_x$ that is nontrivially
pinned by a $z$-independent random potential, that is short-ranged in
the $x-y$ plane.  Standard analysis gives an effective Hamiltonian
\begin{eqnarray}
\tilde H &=& \int_\rv\left[\oh u_x\hat{\tilde\Gamma} u_x 
+ U_0(x,y)\partial_x u_x + U(x,y, u_x(\rv))\right], 
\;\;\;\;\;\;\;\;
\label{Hxx}
\end{eqnarray}
where $\hat{\tilde\Gamma}$ is the effective elastic kernel, whose
Fourier transform $\tilde\Gamma({\bf q})= \left(\Gamma_{xx}\Gamma_{zz}
  - \Gamma_{zx}^2\right)/\Gamma_{zz}=1/(\hat\Gamma^{-1})_{xx}$ is given by
\begin{eqnarray}
\tilde\Gamma(q_x,q_y,q_z)&=& \frac{\left(K q_y^2 + \mu q_\perp^2\right)
\left(K q_y^2 + (2\mu+\lambda) q_\perp^2\right)}
{\left(K q_y^2 + \mu q_\perp^2 + (\mu+\lambda) q_z^2\right)}.\;\;\;\;\;\;
\label{Gamma_eff}
\end{eqnarray}
As expected, though long-ranged, it scales as $q^2$, by power-counting akin to a random
field xy-model.

Minimization of $H$ in \rf{Hfinal} over $u_z$ shows that in the ground
state (relevant for zero temperature), neglecting boundary
contributions and possible random symmetry breaking along $z$
(possible for strong disorder)
\begin{equation}
  \hat\Gamma_{zz}u_z = -(\mu+\lambda)\partial_z\partial_x u_x(x,y)=0.
 \label{uz0}
\end{equation}
Thus, as anticipated, vortex lattice phonons along columnar defects,
$u_z$ experience no disorder-induced distortions or
pinning\cite{commentRandomStress}, fully controlled by
thermally-induced fluctuations. The corresponding two-point correlator
$C_{zz}(\rv)=\oh\overline{\langle\left(u_z(\rv)-u_z(0)\right)^2\rangle}$
is then given by
\begin{eqnarray}
C_{zz}(\rv) &=&
T\int_\qv\frac{\left[K q_y^2 + \mu q_\perp^2 + (\mu+\lambda)
q_x^2\right]\left(1-e^{i\qv\cdot\rv}\right)}
{\left(K q_y^2 + \mu q_\perp^2\right)
\left(K q_y^2 + (2\mu+\lambda) q_\perp^2\right)},
\nonumber\\
_{\rv\rightarrow\infty}&=&u_{z,\rm rms}^2 
\approx\frac{T}{\overline K a},\;\;\;\;\;\;\;\;
\label{Czz}
\end{eqnarray}
where $\overline K$ is an effective stiffness (a function of
$K,\mu,\lambda$) whose detailed form is unimportant, and
$a\sim 2\pi/\Lambda$ is lattice cutoff.  As indicated above, the key
point is that in 3D this integral is convergent at small wavevectors
and thus at long scales the phonon correlator asymptotes to a finite
$\rv$-independent constant, and therefore at low temperature exhibits
a stable periodic (layered) smectic order along columnar defects. We
thus expect at low $T$, the structure function for vortex
configuration transverse to columnar defects,
Fig.\ref{fig:SmVGfig}(a), to exhibit true smectic Bragg peaks for
momentum transfer along along columnar defects ($q_z$), with amplitude
reduced by a Debye-Waller factor, $e^{- Q_z^2 u_{z,\rm rms}^2}$, as
illustrated in Fig.\ref{StructureFncSVGfig}.

As the temperature is raised, at sufficiently high $T$, such that
$u_\text{z,rms}$ increases to a fraction of a lattice constant -- a
Lindemann criterion, we expect that the smectic order will melt at
a temperature,
\begin{equation}
T_\text{melt-Sm}\approx c_L a^3\overline K,
\label{TmeltSm}
\end{equation}
where $c_L$ is a Lindemann number of order $1$ to be fit by
experiments.  At low temperature below $T_\text{melt-Sm}$, SmVG state
is stable and the nontrivial part of the problem reduces to a
pinning-induced distortions of a single phonon, $u_x$, transverse to
columnar defects.

\section{Pinning of transverse phonons}
\label{pinningTransverse}

We now study Hamiltonian \rf{Hxx}, analyzing the statistics of the
transverse phonon $u_x(x,y)$ in the presence of Gaussian short-range
correlated $z$-independent pinning potentials,
$U(\xv, u_x(\xv,z)), U_0(\xv)$, and elastic kernel $\tilde\Gamma(\qv)$,
\rf{Gamma_eff}.

\subsection{Perturbative Larkin analysis}
\label{LarkinSec}
Vortex lattice $u_x(\rv)$ distortions are characterized by a sum of
thermal and random pinning contributions, $C_{xx}(\rv)=C_{xx}^T(\rv) +
C_{xx}^\Delta(\rv)$, where
\begin{eqnarray}
C_{xx}^T(\rv)&=&
\oh\overline{\langle\left(u_x(\rv)-u_x(0)\right)^2\rangle_c},\\
C_{xx}^\Delta(\xv)&=&
\oh\overline{\langle u_x(\rv)-u_x(0)\rangle^2},
\label{CxxTandDelta}
\end{eqnarray}
where subscript $c$ denotes ``connected'' correlation function
$\langle\phi\phi\rangle_c\equiv\langle(\phi -
\langle\phi\rangle)^2\rangle =\langle\phi\phi\rangle -
\langle\phi\rangle\langle\phi\rangle$. Above, in contrast to the
thermal component, $C_{xx}^T(\rv)$, we explicitly indicated
$C_{xx}^\Delta(\xv)$ to be a function of $\xv$, rather than full
$\rv = (\xv,z)$, i.e., independent of $z$. At short scales, such that
$u_x$ phonon distortions are small compared to the lattice constant
and disorder correlation length, we can follow Larkin\cite{Larkin} and
expand the random potential $U(\xv, u_x(\xv))$ in $H$, \rf{Hfinal} or
\rf{Hxx} to linear order in $u_x(\xv)$ about an undistorted state (a
random force approximation equivalent to the Imry-Ma estimate), that
can be then analyzed exactly in this perturbative
regime.\cite{ImryMa,Larkin,DSFisherFRG,GLbragg,RTaerogel}.

At low temperature
$C_{xx}^\Delta(\xv)\equiv G_{xx}^\Delta(0)-G_{xx}^\Delta(\xv)$
dominates, with phonon correlations captured by the propagator
$G_{xx}^\Delta(\xv)=\overline{\langle u_x(\rv)\rangle \langle
  u_x(0)\rangle}$, and mean-squared distortions given by
\begin{eqnarray}
  G_{xx}^\Delta(0)&\approx&
  \int_{\qv_x,q_y}\frac{\Delta}{|\tilde\Gamma(\qv_x,q_y,0)|^2},\\
                  &\approx&\int_{\qv_x,q_y}\frac{\Delta}
                            {\left[K q_y^2 + (2\mu+\lambda)
                            \qv_x^2\right]^2},\nonumber\\
                  &=&\frac{\Delta}{\pi\tilde K^2}\;L^2,\;\;\;\text{for}\; d=3,\\
                  &=&\frac{C_{d-2}\Delta}{\tilde K^2}\;L^{5-d},\;\;\;\text{for}\; d<5,
\label{G_xx03d}
\end{eqnarray}
where $\tilde K^2\equiv\sqrt{K(2\mu+\lambda)^3}$ is the effective
elastic constant, $\Delta\equiv\Delta(0)=-R''(0)$ is the random force
correlator, and in the last equality, for later convenience we
generalized the analysis to $d$ dimensions, denoting $d$-dimensional
coordinate vector $\rv \equiv (\xv, z)$ and $(d-1)$-dimensional
coordinate vector transverse to $\hat{\bf z}$ as
$\xv = (\xv_\perp,y)$. Above, for convenience we chose a cylindrical
momentum cutoff with $-\infty<q_y<\infty$ and defined
$C_d=S_d/(2\pi)^d=2 \pi^{d/2}/\Gamma(d/2)/(2\pi)^d$ ($C_1 = 1/\pi$,
$C_3 = 1/(2\pi^2)$), with $S_d$ a surface area of a $d$-dimensional
unit sphere. In above analysis we also neglected the long-wavelength
disorder $U_0(\xv)$, that is subdominant at scales shorter than
$\xi_\text{L}$, where it gives distortions that scale logarithmically
in $d = 3$, and are finite for $d > 3$.

Above strongly divergent $u_{\text{rms}}$ distortions are expected due
to the correlated $z$-independent random columnar potential,
contrasting with the corresponding point-disorder $L^{4-d}$
growth.\cite{ImryMa,Larkin,DSFisherFRG,GLbragg, RTaerogel}. For
$d < d_{\text{lc}} = 5$ the perturbative result \rf{G_xx03d} predicts
its own breakdown at sufficiently long scales $L > \xi_{\text{L}}$,
where Larkin correlation length (up to constants of $O(1)$) is given
by
\begin{eqnarray}
\xi_{\text{L}}&=& \left(a^2 \tilde K^2/\Delta\right)^{1/(5-d)}.
\label{xiLarkin}
\end{eqnarray}

Through this perturbative analysis, a combination of \rf{Czz} and
\rf{G_xx03d} predicts that in the physical case of three dimensions,
vortex configurations transverse to columnar defects
(Fig.\ref{fig:SmVGfig}(a)) are characterized by a periodic order along
columnar defects, with $u^z_{\text{rms}}\ll a$ at low $T$ and
divergent distortions, $u^x_{\text{rms}}$ perpendicular to columnar
defects, that we will show grow logarithmically on scales longer than
$\xi_{\text{L}}$. Thus, as illustrated in Figs.\ref{fig:SmVGfig}(a,b),
\ref{StructureFncSVGfig}, this 3D transverse vortex state, that we
dubbed a ``smectic vortex glass'' is an array of $y$-directed vortex
sheets confined to the $x-y$ plane, exhibiting long-range periodic order
along the $\zh$-directed columnar defects, with true Bragg peaks along
$q_z$ \rf{StructureFncSVGfig}, and, as we will show below,
quasi-long-range order (characterized by quasi-Bragg peaks) within the
$x-y$ plane.

\subsection{Replicated model}

We now focus on the positional order within the $\xv_\perp-y$ plane
transverse to columnar defects, characterized by distortions
$u(\rv)\equiv u_x(\rv)$, where for simplicity of notation we have
dropped subscript $"x"$.

To compute self-averaging quantities, (e.g., the disorder averaged
free energy and correlation functions) it is convenient (but not
necessary) to employ the replica ``trick''\cite{EA}, which allows us
to work with a translationally invariant field theory at the expense
of introducing $n$ replica fields, with the $n\rightarrow 0$ limit to
be taken at the end of the calculation. For the free energy this
procedure relies on the identity for the $\ln z$ function
\begin{equation}
{\overline F}=-T\overline{\ln Z}=-T\lim_{n\rightarrow
0}\frac{\overline{Z^n}-1}{n}\;.
\label{replica}
\end{equation}
After replicating and integrating over the random potential
$U[u(\rv),\xv]$ using \rf{Rdefine}, we obtain
\begin{equation}
\overline{Z^n}=\int[du_\alpha]e^{-H^{(r)}[u_\alpha(\rv)]/T}\;.
\end{equation}
The effective translationally invariant replicated Hamiltonian
$H^{(r)}[u_\alpha(\rv)]$ is given by
\begin{eqnarray}
H^{(r)}&=&\oh\sum_{\alpha}^n\int_{\xv,z}\left[
u_\alpha\hat{\tilde\Gamma} u_\alpha +\mu_{zx}(\partial_z u_\alpha)^2\right]\nonumber\\
&&-\frac{1}{2T}\sum_{\alpha,\beta}^n\int_{\xv,z,z'}
R[u_\alpha(\xv,z)-u_\beta(\xv,z')],\;\;\;
\label{Hsr}
\end{eqnarray}
where we added an additional elastic term, characterized by a shear
modulus $\mu_{zx}$, that we expect to be generated under coarse-graining. We
will use this Hamiltonian \rf{Hsr} in our subsequent RG analysis of
the system.

\subsection{Physics beyond Larkin scale: functional RG}

As we saw in the Larkin analysis of Sec.\ref{LarkinSec}, at scales
shorter than $\xi_\text{L}$, the transverse phonon distortions $u$ are
small compared to the lattice constant and disorder correlation
length, and thus have been treated
perturbatively\cite{ImryMa,Larkin,DSFisherFRG,GLbragg, RTaerogel}. The
growth of $u_{\text{rms}}\sim L^{(5-d)/2}$ in \rf{G_xx03d} is
indicative of highly nonlinear, nonperturbative effects of disorder at
scales beyond $\xi_\text{L}$, requiring a functional renormalization
group (FRG) treatment\cite{DSFisherFRG}, that can be controlled in an
expansion in $\epsilon=5-d$ about the lower-critical dimension,
$d_{lc}=5$.

To this end it is convenient to work with the translationally
invariant replicated Hamiltonian, $H^\re$, \rf{Hsr}. We employ the
standard momentum-shell RG transformation\cite{WilsonKogut,DSFisherFRG},
by separating the phonon field into a long and short scale
contributions according to
$u_\alpha(\rv)=u^<_\alpha(\rv) + u^>_\beta(\rv)$ and perturbatively in
nonlinearity $R[u^>_\alpha(\xv,z)-u^>_\beta(\xv,z')]$ integrate out
the high wavevector components $u^>_\alpha(\rv)$, that take support in
an infinitesimal shell $\Lambda/b < q < \Lambda\equiv 1/a$, with
$b=e^{\delta\ell}$. We follow this with a rescaling of lengths and the
long wavelength part of the field
\begin{eqnarray}
\xv&=&b\, \xv',\label{xb}\\
z&=&b^\omega\, z',\label{zb}\\
u^<(b\,\xv', b^\omega z')&=&u'(\xv',z),\label{ub}
\end{eqnarray}
so as to restore the UV cutoff back to $\Lambda$. Based on disorder
correlated along $z$ we anticipated a nontrivial anisotropy between
$\xv$ and $z$, encoded in $\omega\approx 1 + O(\epsilon)$. In
\rfs{ub} we made a convenient choice of a zero scaling dimension for
the real-space displacement field $u(\xv,z)$. This is dictated by the
convenience of keeping the periodicity of the disorder variance $R(u)$
fixed at $2\pi/Q$.

The above rescaling leads to zeroth order RG flows of the effective
elastic constant $\tilde K$ (that represents $K,\mu,\lambda$),
$\mu_{zx}$ and disorder pinning potential $R(u)$, that are given by
\begin{eqnarray}
  \tilde K(b)&=&b^{d-3+\omega}\tilde K\;,\label{Kflow0}\\
  \mu_{zx}(b)&=&b^{d-1-\omega}\mu_{zx}\;,\label{muzflow0}\\
  R(u,b)&=&b^{d-1+2\omega}R(u),\label{Rflow0}
\label{KRflow0}
\end{eqnarray}
indicating that in $d>2$ the effective strengths of both elastic and
pinning energies grow at long scales relative to the thermal energy,
$T$. This is a reflection that in $d>2$ the physics is controlled by
the zero-temperature ground-state competition between elastic and
pinning energies, at long scales both much larger than the thermal
energy.  Equivalently, to emphasize this physics we can rescale $T$
according to
\begin{eqnarray}
T(b)&=&b^{-(d-3+\omega)}T,\\
    &\equiv& b^{-\Theta}T,
             \label{Tflow}
\end{eqnarray}
with $\Theta\approx d-2 +O(\epsilon)$, so as to keep the elastic
energy fixed at order $1$.  With this convenient rescaling convention,
the measure of the effective pinning strength grows according to
\begin{equation}
R(u,b)=b^{5-d}R(u),\label{RT2flow0}
\end{equation}
modified by a factor $(T(b)/T)^2=b^{-2\Theta}$ relative to that in
\rfs{Rflow0} due to the factor of $1/T^2$ in $H^\re/T$,
\rf{Hsr}. Equivalently, without the rescaling of $T$, the
dimensionless combination that arises in the coarse-graining analysis
is given by $R(u)/\tilde K^2$, and its zeroth order flow is given by
\rfs{RT2flow0}.

In either convention we find that for $d<d_{lc} = 5$, the Gaussian
disorder-free fixed point is unstable, indicating that the influence
of random pinning grows at long scales relative to the elastic energy,
consistent with the scaling and Larkin analysis above.

The statistical symmetry\cite{TonerDiVincenzo,DSFisherFRG,GLbragg,
  RTaerogel} of the bulk Hamiltonian, $H$ \rf{Hfinal}, under an
arbitrary local distortion, $u(\xv,z)\rightarrow u(\xv,z) + \chi(\xv)$
guarantees that the flow of $\tilde K(b)$, \rf{Kflow0} and,
equivalently, the thermal exponent
\begin{equation}
  \Theta=d-3+\omega\approx d-2 + O(\epsilon),
  \label{ThetaExact}
\end{equation}
are {\em exact}, i.e., do not experience any coarse-graining
corrections. This can equivalently be seen from the replicated
Hamiltonian \rf{Hsr}, where the pinning nonlinearity,
$R[u_\alpha(\xv,z)-u_\beta(\xv,z')]$ depends only on the difference
between fields at distinct $z$ coordinates and replicas, i.e.,
independent of the ``center of mass'' field
$\sum_{\alpha=1}^n u_\alpha(\xv)$. That is, the only nonlinearity in
$H^\re$ exhibits a symmetry of a replica- and $z$-independent local
distortion $u_\alpha(\xv,z)\rightarrow u_\alpha(\xv,z) + \chi(\xv)$
and under coarse-graining can therefore only generate terms that also
exhibit this symmetry. Thus, it cannot generate a correction to the
elastic term that clearly lacks this symmetry, implying that
$\tilde K$ is {\em not} renormalized by the pinning disorder. In
contrast, as we discuss below, the shear modulus $\mu_{zx}$ indeed is
strongly renormalized by the pinning potential under coarse-graining.

\subsubsection{Pinning}
An important consequence of the periodic nonlinearity $R(u)$ and the
effective zero-temperature physics, first emphasized by
D.S. Fisher\cite{DSFisherFRG} is that all monomials or (equivalently)
harmonics in the expansion of $R(u)$ are equally relevant in
$2 - O(\epsilon) < d < d_{lc}=5$. Thus, a {\em functional} RG analysis that follows the
coarse-graining flow of the whole {\em function} $R(u,b)$ is
necessary. The method is by now quite standard\cite{DSFisherFRG,GLbragg,
  RTaerogel} and is straightforwardly adapted to the correlated
pinning problem\cite{BalentsEPL}, characterized by $H^\re$, \rfs{Hsr}.

We limit the FRG analysis to one-loop order, performing the
momentum-shell integration over the high-wavevector components
$u^>_\alpha$ perturbatively in the nonlinearity
$H_p = -\frac{1}{2T}\sum_{\alpha,\beta}^n\int_{\xv,z,z'}
R[u_\alpha(\xv,z)-u_\beta(\xv,z')]$, and controlled by an
$\epsilon = 5-d$ expansion. The correction to the Hamiltonian due to
the coarse-graining is given by
\begin{equation}
\delta H^\re[u^<_\alpha]=\langle
H_{p}[u^<_\alpha + u^>_\alpha]\rangle_>
-\frac{1}{2T}\langle H_{p}^2[u^<_\alpha + u^>_\alpha]\rangle_>^c\ldots\;,
\label{deltaHsr}
\end{equation}
where the averages over short scale fields, $u^>_\alpha$ above are
performed with the quadratic (elastic $\tilde K$) part of $H^\re$. The
superscript $c$ denotes a cumulant average,
$\langle H_p^2\rangle^c =\langle H_p^2\rangle - \langle H_p\rangle^2$.

%
%
%

Based on Eqs.\rf{Tflow},\rf{ThetaExact}, temperature is (dangerously)
irrelevant for $d > 2$, and we thus focus on the zero temperature
limit, in which the lowest order contribution comes at a second order
in $R(u)$, given by
\begin{widetext}
%
\begin{eqnarray}
\delta H^\re&\approx&
                      -\frac{1}{16 T^3}\sum_{\alpha_1,\beta_1,\alpha_2,\beta_2}
                      \int_{\substack{\xv_1,z_1,z'_1\\\xv_2,z_2,z'_2}}
R''[u^<_{\alpha_1}(\xv_1,z_1)-u^<_{\beta_1}(\xv_1,z'_1)]
R''[u^<_{\alpha_2}(\xv_2,z_2)-u^<_{\beta_2}(\xv_2,z'_2)]
  I^{\alpha_2\beta_2}_{\alpha_1\beta_1}(\xv_1,z_1,z'_1;\xv_2,z_2,z'_2)\;,\nonumber\\
\label{deltaH2}
\end{eqnarray}
where in above, the prime indicates a partial derivative with respect to 
$u$, and
\begin{eqnarray}
I^{\alpha_2\beta_2}_{\alpha_1\beta_1}
&=&\oh\langle (u_{\alpha_1}^>(\xv_1,z_1)-u_{\beta_1}^>(\xv_1,z'_1))^2
(u_{\alpha_2}^>(\xv_2,z_2)-u_{\beta_2}^>(\xv_2,z'_2))^2\rangle_>^c
\;,\label{Iab1}\nonumber\\
&=&8\left[\oh\delta_{\alpha_1\alpha_2}\delta_{\beta_1\beta_2}
G^>_T(\xv_1-\xv_2,z_1-z_2) G^>_T(\xv_1-\xv_2,z'_1-z'_2)
-\delta_{\alpha_1\alpha_2}\delta_{\alpha_1\beta_2}
    G^>_T(\xv_1-\xv_2,z_1-z_2) G^>_T(\xv_1-\xv_2,z_1-z'_2)\right]^2.
\label{Iab2}\nonumber\\
  \label{Iab}
\end{eqnarray}
\end{widetext} 
Above we kept only the most relevant two-replica thermal terms, with a
thermal (zero-disorder) momentum-shell propagator,
$G^>_T(\xv,z)=\langle u^>(\xv,z) u^>(0,0)\rangle^>_0$.  Using the
short-range property of this propagator and comparing to $H_p$, we
then obtain
\begin{equation}
\delta R(u)\approx \delta\ell g_2
\left(\oh R''(u)R''(u)-R''(u)R''(0)\right)
\label{deltaR2}\;,\ \ \ \ 
\end{equation}
where the constant $g_2$ is defined by
\begin{eqnarray}
  g_2\,\delta\ell&=&T^{-2}\int_{\delta\xv,\delta z,\delta z'}
 G^>_T(\delta\xv,\delta z)  G^>_T(\delta\xv,\delta z') \nonumber\\
&=& \int^>_{\qv_x,q_y}\frac{1}{|\tilde\Gamma(\qv_x,q_y,0)|^2}, \nonumber\\
&=&\int_{\Lambda e^{-\delta\ell}}^\Lambda
\frac{d^{d-2} q_x}{(2\pi)^{d-2}}\int_{-\infty}^\infty \frac{d q_y}{2\pi}
\frac{1}{\left[K q_y^2 + (2\mu+\lambda) \qv_x^2\right]^2},\nonumber\\
&=&\frac{C_{d-2}\Lambda^{d-5}}{\sqrt{K(2\mu+\lambda)^3}}\,\delta\ell\;.
    \label{g2}
\end{eqnarray}

Combining this with the rescalings Eq.~(\ref{Kflow0}-\ref{Rflow0}),
we obtain the FRG flow equation for a dimensionless measure of pinning
disorder, $\hR(u)\equiv g_2R(u)$,
\begin{eqnarray}
\partial_\ell\hR(u)&=&\epsilon\hR(u) 
+ \frac{1}{2} \hR''(u) \hR''(u)-\hR''(u)\hR''(0).\nonumber\\
\label{FRGflowR}
\end{eqnarray}
We note that aside from constant prefactors ($g_2$) in the definition
of $\hR(u,\ell)$ and the raised (from $4$ to $5$) lower-critical
dimension giving $\epsilon=5-d$, this FRG flow is identical to that of
the uncorrelated pinning disorder\cite{DSFisherFRG,GLbragg, RTaerogel},
with the fixed point function for the random force correlator given by
\begin{equation}
\hat{\Delta}_*(u) = -\hR''_*(u)=\frac{\epsilon}{6Q^2}
\left[(Q u-\pi)^2-\frac{\pi^2}{3}\right], 
\label{fixedpointRpp}
\end{equation}
periodically extended with period $2\pi/Q$.

\subsubsection{Breakdown of elasticity}
As noted above, elastic moduli $K,\mu,\lambda$ characterizing
$z$-independent distortions $u(\xv,y)$ are protected by the $\xv-y$
plane statistical symmetry, and thus remain fixed under
coarse-graining. In contrast, it is straightforward to show that the
$\mu_{zx}$ modulus for $\partial_z u$ shear is strongly enhanced by
pinning, acquiring a correction
\begin{equation}
  \delta\mu_{zx} \approx -g_2 \Delta''(0)\mu_{zx}\delta\ell ,
\end{equation}
that in the absence of point disorder and at zero temperature is
divergent since $\Delta(u)$, \rf{fixedpointRpp} exhibits slope
discontinuity at $u=0$.

To understand the physical implications of this result requires a
careful analysis of the dangerously irrelevant temperature and/or
point disorder, neglected so far. These smoothen the $u=0$ singularity
within the boundary layer that shrinks on coarse-graining with
$\ell$.  Conveniently, such analysis has been carefully done in
Ref.~\onlinecite{BalentsEPL} for a physically distinct, but mathematically
related 'toy' model of a Bose glass.  The upshot is that pinning
generates a nonanalytic term,
\begin{equation}
  \delta H = \oh\sigma_c\int_\xv |\partial_z u|,
\end{equation}
arising from the singularity $\Delta(u)\sim -|u| + const.$ around
$u=0$ outside the boundary layer. As discussed in
Ref.~\onlinecite{BalentsEPL}, this results in a finite threshold
$\sigma_c$ response to an external shear stress $\sigma_{zx}$, with
coupling $\sigma_{zx}\partial_z u$. This is the SmVG's counterpart of
the ``transverse Meissner effect'' in the Bose glass geometry (with
field oriented along columnar defects)\cite{NelsonVinokur}. However,
in contrast, here it is unclear to us how to probe this novel,
effectively divergent $\mu_{zx}$, characteristic of the SmVG.

\subsubsection{Correlation function}

With this RG analysis in hand, we can now calculate the transverse
correlation function $C_{xx}^\Delta(\xv)$, \rf{CxxTandDelta}, that
dominates the asymptotics of vortex line distortions at scales beyond
the perturbative Larkin scale.  To this end we first examine the
corresponding momentum correlation function,
\begin{equation}
  C({\qv})\equiv\frac{\overline{\langle u(\qv)\rangle\langle u(\qv')\rangle}}
    {\delta^d(\qv +\qv')\delta(q_z)}
\end{equation}
To compute this function we utilize the standard RG matching
analysis\cite{GLbragg,RTaerogel}, that allows us to relate a
correlation function at a small wavevector $\qv_x$ of our interest
(which is impossible to calculate in perturbation theory due to the
infra-red divergences) to the same correlation function at a large
wavevector, $b \qv_x$, which can be straightforwardly calculated in a
controlled perturbation theory,
\begin{equation}
C(\qv_x,q_y,0)= b^{d - 1} C(b\qv_x, b q_y, 0, K, \mu,\lambda,
R(u,b))
\label{Cmatch1}\;.
\end{equation}
Choosing $b\qv_x =\Lambda$ for small $\qv_x$, such that $b$ is large
to take $R(u,b\gg 1)\rightarrow R_*(u)$ to its fixed point
\rf{fixedpointRpp}, and computing the right hand side perturbatively
in disorder, we find
\begin{equation}
C(\qv_x,q_y,0)= \frac{1}{q_x^{d - 1}}\frac{\hat\Delta_*(0)\Lambda^{d-5}/g_2}
{\left[K(q_y/q_x)^2+ (2\mu +\lambda)\right]^2}
\label{Cmatch2}\;.
\end{equation}
In coordinate space, at long scales beyond $\xi_L$, we find (for equal
$y$ separation),
\begin{equation}
C(\xv)=\oh\overline{\langle u(\xv)-u(0)\rangle^2}\approx Q^{-2}
\eta_d\ln(x/a),
\label{CxxLog}
\end{equation}
for all dimensions $d$ in the range $2-O(\epsilon) < d < 5$, where $A$
is a universal amplitude that to one loop order is given
\begin{eqnarray}
  \eta_d = \frac{\pi^2\epsilon}{9},
\end{eqnarray}
identical to that found in Refs.~\onlinecite{GLbragg,BalentsEPL}.

These translational correlations are best probed by the structure
function, a Fourier transform of the density-density correlation
function,
\begin{equation}
S(\qv)=\sum_{\rv^\perp_n} e^{-i{\bf q}\cdot \rv^\perp_n}
\overline{\langle e^{i \qv\cdot(\uv(\rv^\perp_n)-\uv(0))}\rangle}.
\label{SqDefn}
\end{equation}

For scattering along columnar defects, $\qv = q_z\zh$ and using finite
$u_z$ distortions set by $u_{\text{z,rms}}$, \rfs{Czz}, we find
``true'' $\delta$-function Bragg peaks, suppressed by the thermal
Debye-Waller factor, $I_{Q_z}= e^{-Q_z^2 T/(\overline{K} a)}$, as
advertised in Eq.\rf{SqzBragg}.

In contrast, for scattering at a wavevector $\qv = q_x\xh$, transverse
to columnar defects and the applied field (vortex lines), we utilize
the Gaussian approximation (which amounts to ignoring higher order
cumulants) and the $u_x-u_x$ correlation function computed above,
\rf{CxxLog}. We thereby find
\begin{eqnarray}
 S(q_x)&=&\sum_{x_n} e^{-i q_x x_n}e^{- q_x^2 C(x_n)}\;,\nonumber\\
        &\approx&\sum_{x_n} e^{-i q_x x_n}\frac{1}{|x_n|^{\eta_{q_x}}}\;,\nonumber\\
&\sim& \sum_p \frac{1}{|q_x - p Q_x|^{1-p^2\eta}},            
  \label{SqxQuasiBragg}
\end{eqnarray}
where the universal power-law exponent in 3D, 
$\eta=\eta_3\approx 2\pi^2/9$, and we utilized the Poisson summation
formula to perform $x_n$ summation.  We note that if this approximate
value of $\eta$ is indeed greater than $1$, only cusp singularities,
rather than divergent quasi-Bragg peaks will appear at the reciprocal
lattice vectors, $Q_x$.


\section{Dislocations}

As with other pinning problems\cite{DSFisherFRG,GLbragg,
  DSFisherBG,RTaerogel, stabilityCarpentierBraggGlass1996,
  stabilityKierfeldBraggGlass1997, DSFisherBG}, above analysis
notwithstanding, a vexing problem is that of topological
defects. Namely, it is important to assess the stability of the SmVG
phase to a proliferation of dislocations. For strong disorder, at high
temperatures, and/or in the presence of point disorder, these will
undoubtedly proliferate, destroying topologically-ordered (Bragg
glass) character of the SmVG.  However, at weak disorder and low
temperatures, one can contemplate that dislocation-free Bragg glass
order\cite{GLbragg, DSFisherBG,RTaerogel} will persist to arbitrary
scales of the SmVG state.  Otherwise, our above results will extend
only out to a dislocation scale, $\xi_D$ (set by the average spacing
of proliferated dislocations), that will be long in the limit of weak
disorder and low temperatures, thereby allowing a broad range of SmVG
regime.

In addition to these generic considerations\cite{GLbragg,
  DSFisherBG,RTaerogel}, that is extremely challenging to rigorously
assess, we believe considerations of SmVG are much more favorable for
its stability against dislocations. It is made so by the translational
invariance of disorder along columnar defects ($z$-axis) and vortices
running transverse to them.  This is based on the fact that for SmVG,
(i) \rfs{uz0} strictly gives $u_z = 0$ in the ground state, (ii)
convergent thermal $u_{z,rms}$ fluctuations, that can be made
arbitrarily small at low $T$ according to \rf{Czz}, and (iii) all
observables are expected to be $z$ independent in the ground state
(neglecting a possibility of spontaneous inhomogeneous symmetry
breaking along columnar defects at strong disorder).

To examine these arguments in more detail, we note that a potential
dislocation density in SmVG is defined by a nonsingle-valued
displacement field,
\begin{equation}
  \hat{\bf y}\cdot\grad\times\grad u_i = b_i(\rv)
  \label{dislocation}
\end{equation}
A key consequence of the property (iii), the $z$-independence of
vortex distortions, $u_i(\xv)$, is that with gradients in
\rf{dislocation}, acting within the $x-z$ plane, immediately implies a
vanishing of dislocation density, $b_i(\rv) = 0$.

In addition, we can focus on vortex line order within the $x-y$
planes, with vortices running along the $y$-axis, as illustrated in
Fig.\ref{fig:SmVGfig}(a). Based on properties (i) and (ii), at low
temperature and weak disorder, we can neglect vortex distortions along
columnar defects, taking $u_z = 0$. We then note that with these
conditions, within the $x-y$ planes, the system reduces to an array of
``$1+1$'' dimensional systems of Fisher's vortex
glass\cite{MatthewFisherVG}.  Based on this, it is obvious that
magnetic flux conservation (vortex line continuity) immediately
forbids dislocations (partial lines) within these $x-y$ vortex planes.
This can be seen mathematically by observing that components of the
magnetic flux are generically given by
${\bf B} = B_0(\partial_y u_x, -\grad\cdot{\bf u}, \partial_y u_z)$,
and for $u_z\approx 0$ reduce to
\begin{eqnarray}
  B_x = B_0\partial_y u_x\;,\;\;\;   B_y = -B_0\partial_x u_x\;,
 \end{eqnarray}  
 
 Thus, the vanishing divergence of flux density (absence of monopoles)
 $\grad\cdot{\bf B} = 0$, reduces to,
\begin{eqnarray}
  \partial_x\partial_y u_x - \partial_y\partial_x u_x =0\;,
 \end{eqnarray}  
 which guarantees a single-valued form of the in-plane displacement
 $u_x$, and thus absence of in-plane dislocations.

 Based on the above discussion, we thus argue that at low temperature
 and weak disorder (certainly in the absence of point disorder, though
 may even survive with weak point disorder\cite{DSFisherBG}) the SmVG
 phase remains dislocation-free, topologically-ordered smectic Bragg
 glass\cite{RTaerogel}.


\section{Summary and Conclusions}
\label{conclusion}

Motivated by old experiments on vortex matter tilted away from
columnar defects\cite{JaegerPRL2000PRB2001}, in this paper we studied
a new smectic glass state of vortex matter.  We showed that it arises
for large tilt angles of the magnetic field relative to columnar
defects, at which a Bose glass undergoes a phase transition to this
SmVG.  As summarized in the Introduction, this novel state is
characterized by a periodic translational order along the columnar
defects (exhibiting true Bragg peaks) and Bragg
glass\cite{GLbragg,DSFisherBG,RTaerogel} logarithmically rough vortex
distortions transverse to columnar defects, that lead to quasi-Bragg
power-law peaks in the structure function.  SmVG is also characterized
by a divergent shear modulus in the plane transverse to the applied
magnetic field, a counter-part of the ``transverse'' Meissner effect
predicted for the Bose vortex glass.\cite{NelsonVinokur,BalentsEPL} We
also predict that SmVG exhibits an infinite electrical transport
anisotropy with a {\em nonzero} longitudinal resistivity transverse to
columnar defects and to the applied magnetic field and a {\em vanishing}
resistivity along the columnar defects.

In this manuscript, we focussed on the purely transverse (tilt angle
of $\pi/2$) geometry, but expect that the SmVG phase exhibits a finite
range of tilt angle stability. However, the symmetry of the state
clearly changes away from this purely transverse orientation. We leave
the analysis of a more general geometry, as well as the nature of the
Bose-to-smectic vortex glass phase transition to future studies.

\begin{acknowledgments}
  It is a pleasure to thank Heinrich Jaeger and Tom Rosenbaum for
  sharing their data with me prior to publication and to acknowledge
  illuminating discussions with John Toner and Pierre LeDoussal.  This
  research was supported by the Simons Investigator Award from the
  James Simons Foundation, and in part by the National Science
  Foundation through the Soft Materials Research Center under NSF
  MRSEC Grant and through the KITP under Grant No. NSF PHY-1748958.  I
  also thank the KITP for its hospitality during my stay as part of a
  sabbatical and the Synthetic Quantum Matter workshop, when part of
  this work was completed.
\end{acknowledgments}



\begin{thebibliography}{6}
\bibitem{GLelastic} Elastic moduli $K$ and $B$ can be derived 
from the Ginzburg-Landau theory and for finite-range interactions (set
by the London penetration length $\lambda$) are wavevector dependent
on scales shorter than $\lambda$, and become dispersionless only 
on scales longer than $\lambda$, i.e., for $n_0\lambda\ll 1$.
For simplicity we will ignore this inessential complication.
\bibitem{commentLatticeAnisotropy} In the presence of columnar defects
  with transversely oriented vortex lattice, generically the
  elasticity will exhibit a biaxial (anisotropic $C_2$ symmetry) form,
  $H_{el} = \frac{1}{2}\int_\rv\left[K_i(\partial_y u_i)^2 +
    c_{xx}u_{xx}^2 + c_{zz}u_{zz}^2 + c_{xz}u_{xx}u_{zz} +
    \mu_{zx}u_{xz}^2\right]$. However, because this does not introduce
  any new qualitative features but technically complicates the
  analysis, we will limit our treatment to isotropic elasticity,
  \rfs{Hel}, except for the divergent shear modulus $\mu_{zx}$ induced
  by columnar disorder.
\end{thebibliography}
\end{document}